\documentclass[submission,copyright,creativecommons]{eptcs}

\providecommand{\event}{FMAS 2021}

\title{Complete Agent-driven Model-based System Testing for Autonomous Systems\thanks{Kerstin Eder was supported in part by the ``UKRI Trustworthy Autonomous Systems Node in Functionality''  under grant number EP/V026518/1. Wen-ling Huang and Jan Peleska were supported in part by the German Ministry of Economics, Project
``HiDyVe -- Highly Dynamic Virtual and Hybrid Validation and Verification'' under  grant agreement~20X1908E.}}
\author{Kerstin I.~Eder\institute{Department of Computer Science, University of Bristol, United Kingdom}
  \email{Kerstin.Eder@bristol.ac.uk} \and    Wen-ling Huang\institute{Department of Mathematics \& Computer Science, University of Bremen, Germany}
  \email{huang@uni-bremen.de} \and Jan Peleska\institute{Department of Mathematics \& Computer Science, University of Bremen, Germany}
  \email{peleska@uni-bremen.de}}

\usepackage{amssymb}
\usepackage{amsmath}
\usepackage{amsfonts}

\usepackage{moreverb}
\usepackage{graphicx}

\usepackage{todonotes}
\usepackage{url}
\usepackage{changepage}
\usepackage[numbers,sort]{natbib}
\usepackage{xspace,boxedminipage}
\usepackage{thumbpdf}
\usepackage{hyperref}
\usepackage{xcolor}
\hypersetup{
    colorlinks,
    linkcolor={red!50!black},
    citecolor={blue!50!black},
    urlcolor={blue!80!black}
}

\usepackage{float}

\usepackage[T1]{fontenc}

\newcommand{\xstop}{x_\text{Stop}}
\newcommand{\tstop}{\Delta_\text{Stop}}
\newcommand{\cmin}{c_\text{Min}}
\newcommand{\vsafe}{v_\text{Safe}}
\newcommand{\vmax}{v_\text{Max}}
\newcommand{\vmin}{v_\text{Min}}
\newcommand{\vconst}{v_\text{const}}

\newcommand{\ul}{\underline}

\newcounter{examplectr}
\newenvironment{example}[1]
{
{\refstepcounter{examplectr}

\medskip
\noindent
\bf Example~\theexamplectr.\label{#1}}
}
{
\unskip\nobreak\hfil\penalty50
      \hskip2em\hbox{}\nobreak\hfil$\Box$%
      \parfillskip=0pt \finalhyphendemerits=0 \par
      
\medskip      
}

\newcommand{\tg}{\mathbf{G}}

\def\titlerunning{System Testing for Autonomous Systems}
\def\authorrunning{Eder, Huang, and Peleska}

\begin{document}
\maketitle

\begin{abstract}
In this position paper, a novel approach to testing complex autonomous transportation systems (ATS) in the 
automotive, avionic, and railway domains is described. It is intended to mitigate  some of
the most critical problems regarding verification and validation (V\&V) effort for ATS. 
V\&V is known to become infeasible for complex ATS, when using conventional methods only. The approach advocated here uses complete testing methods on the module level, because these establish formal   proofs
 for the logical correctness of the software. Having established 
logical correctness, system-level tests are performed  in simulated cloud environments and on the target system. To give evidence that ``sufficiently many'' system tests have been performed with the target system, a formally justified coverage criterion is introduced. To optimise the execution of very large system test suites,  we advocate an online testing approach where multiple tests are executed in parallel, and test steps are identified on-the-fly. The coordination and optimisation of these executions is achieved by an agent-based approach. Each aspect of the testing approach advocated here is shown to either be consistent with existing standards for development and V\&V of safety-critical transportation systems, or it is justified why
it should become acceptable in future revisions of the applicable standards. 
\end{abstract}
% ==============================================================================================
\section{Introduction}\label{sec:intro}

\subsubsection*{Motivation}
Autonomous transportation systems (ATS) in the automotive, avionic, or railway domains are highly complex and at the same time safety-critical. It is a widely accepted belief that the verification and validation (V\&V)  effort   (including  the test effort) for   assuring an acceptable degree of 
safety and reliability in complex ATS
will become so high   that it can no longer be performed with conventional methods~\cite{RR-1478-RC}. In particular, it cannot be expected that all the necessary system tests will be executable on the integrated target system (vehicle, train, or aircraft). Instead, a major portion of the tests needs to be executed  concurrently in  simulation environments. This approach, however, needs special attention from the perspective of applicable safety-related standards and certification rules: for good reasons, it has to be justified why simulation environments are sufficiently trustworthy ``replicas'' of the real target systems and their operational environments, so that certification credit can be obtained for these tests, though they have not been executed with the original equipment and the real operational environment.

% ----------------------------------------------------------------------------------------------
\subsubsection*{Objectives}

This contribution is a position paper: we outline a novel approach to testing complex ATS
and justify each building block of this approach by references to existing theories from the field of formal methods or motivate them at least by means of illustrating examples. The novelty consists in a new combination of existing theories and technologies, and in a careful consideration of applicable standards and pre-standards in the automotive, railway, and avionic domains~\cite{iso26262-4,iso21448,CENELEC50128,DO178C}. Specifically, our approach is as follows.

\noindent
(1) It is proposed to combine so-called \emph{complete} software test strategies on the module level with scenario-based system tests. A test suite generated according to a specific strategy is complete, if it guarantees under certain hypotheses that every correct implementation will pass all test cases and every faulty implementation will fail at least one test case. Correctness is either defined by means of a conformance relation (refinement,   equivalence, or variants thereof) to a given reference model, or by means of a set of property specifications to be fulfilled by the implementation. Here, the model-based approach is used. On the system level, test scenarios are created from more comprehensive system models whose behavioural semantics can be represented by symbolic finite state machines (SFSM)~\cite{DBLP:conf/icst/Petrenko16}, extended by control state invariants for time-continuous and discrete variables. This extension of SFSMs can be interpreted as a restricted variant of hybrid automata, as introduced in~\cite{Hen96}.
The system-level models can be traced back to module-level models, and this relationship can be exploited to obtain meaningful coverage values for system tests.

\noindent
(2) On the system level, tests are performed 
concurrently, following the \emph{online testing} paradigm~\cite{DBLP:conf/fates/LarsenMN04},  
where input data to the \emph{system under test (SUT)} are calculated on the fly from a system model for each test step which is  part of a test case. Also, the SUT's reactions are checked in real-time against the system model. A portion of these concurrent system tests will be performed using the original equipment in the real operational environment, while the rest is executed in cloud-based simulation environments. To coordinate this concurrent effort, an agent-based approach is used -- we use the term \emph{agent-based system testing (ABST)}. 
As pointed out in~\cite{9282616}, the main advantage  of using agents in testing is their ability to perform
autonomous actions. 
They can decide to ``push'' test executions into specific directions, pursuing different goals, such as 
coverage maximisation, prioritisation, or investigation of critical functional aspects of the SUT.

%%%\emph{multi-agent system (MAS)}

%%%\emph{belief-desire-intention (BDI)} agents

% ----------------------------------------------------------------------------------------------
\subsubsection*{Main Contributions}

To the best of our knowledge, the approach discussed in this paper considers the following aspects for the first time.
(1)  The combination of complete testing strategies on the module level with complementary system tests whose degree of completeness can also be measured;
(2) the agent-based approach to maximise system test coverage during online testing;
(3) the investigation of the 
impact of applicable existing and future standards on the admissibility of cloud-based tests
for the purpose of achieving certification credit;
(4) the exploitation of test models created during module testing for the purpose of system test coverage assessment.

% ----------------------------------------------------------------------------------------------
\subsubsection*{Overview}

In Section~\ref{sec:train}, an example is presented, modelling an autonomous freight train controller. This example will be used in subsequent sections
 to illustrate the aspects of the comprehensive test approach advocated in this paper.
In Section~\ref{sec:module}, we discuss complete test methods and their applicability to module tests in the cloud.
We present our approach to agent-based system testing in the cloud and on the original target systems in Section~\ref{sec:systemtest} and show how results from module testing can be used to calculate coverage.
In each of these sections, the certification-related aspects are discussed where appropriate.
Conclusions and plans for future work are presented in Section~\ref{sec:conc}.

Throughout the paper, we refer to related work where appropriate. The technical report~\cite{eder_kerstin_2021_5203111}    contains a comprehensive section on other work related to the approach advocated here. Moreover, this technical report 
contains the full model of the autonomous train controller presented in Section~\ref{sec:train}.

% ==============================================================================================
\section{Running Example -- Autonomous Freight Train Control System}\label{sec:train}

% ....................................................................................
\subsubsection*{System Description}

\begin{figure}[h]
%%\hspace*{-40mm}
\begin{center}
\includegraphics[width=\textwidth]{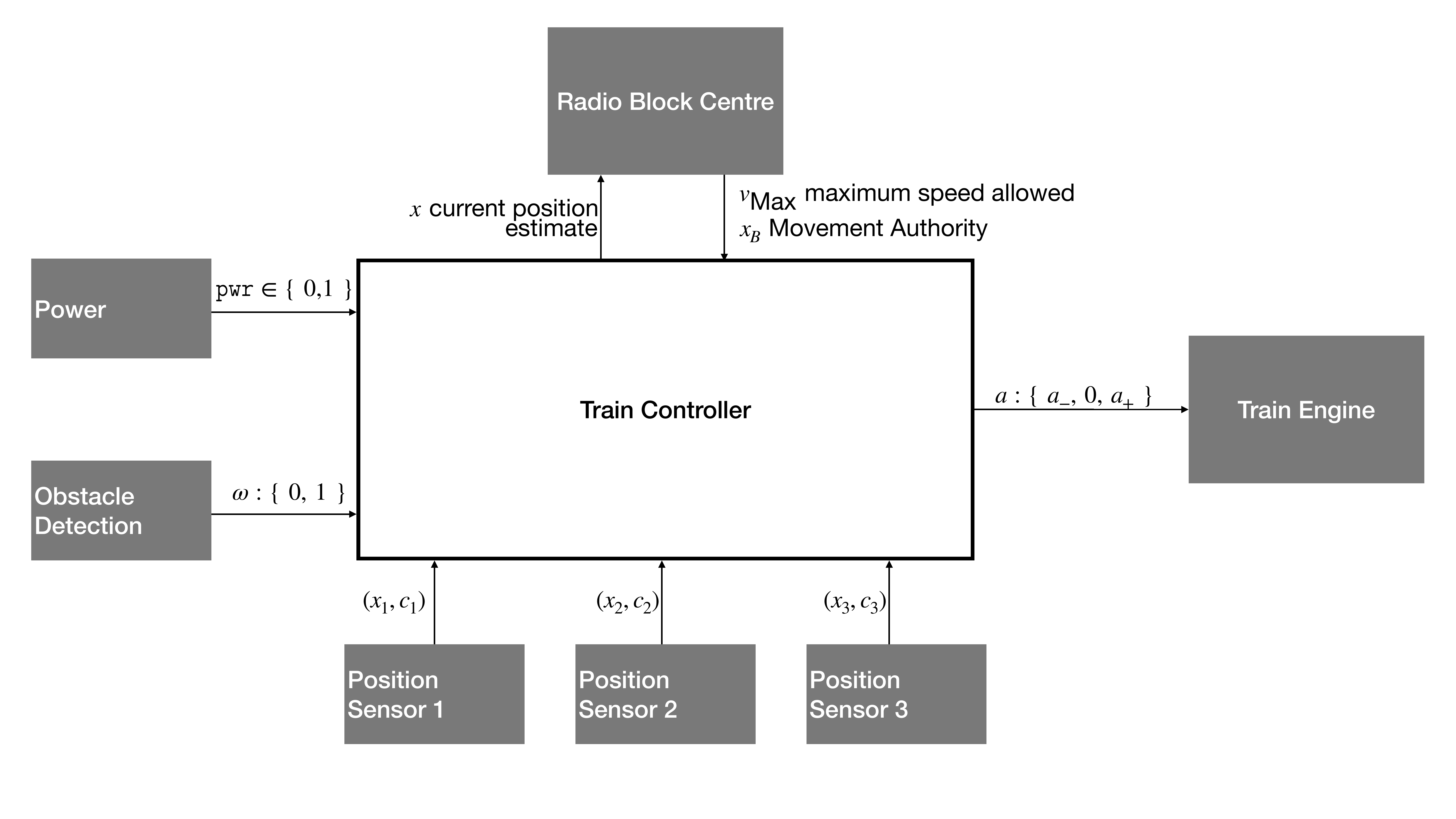}
\end{center}
%%\vspace*{-10mm}
\caption{Train controller, location sensors, and engine interface.}
\label{fig:trainctrl}
\end{figure}

Consider a control system for an autonomous freight train, with interfaces as depicted in Fig.~\ref{fig:trainctrl}. The controller is only active when powered (${\tt pwr} = 1$).
Resetting the controller is performed by switching the power off and on again. 
%For limiting the size of this example, it is assumed that the train is always halted when the power is switched on (no resets while the train is driving).
The controller acts on the train engine with a simplified interface $a$ carrying three commands $a_-$ (negative acceleration,  brake the train), $0$ (no acceleration, keep current velocity -- this state is called \emph{coasting}), and $a_+$ (accelerate the train). For the sake of simplicity, only one deceleration and one acceleration value is considered. The decisions about braking or accelerating depend on several inputs. The \emph{radio block centre (RBC)} sends a \emph{movement authority (MA)} to proceed up to track coordinate $x_B$ which is greater than the train's starting position $x_A$. The train is expected to proceed until $x_B$ and stop there.\footnote{This part is slightly simplified: according to the ETCS standard~\cite{ETCS}, new movement authorities $x_B' > x_B$ may be received while the train is driving, so that a stop at $x_B$ is not required.} Conversely, the train transmits its current position estimate $x$ to the RBC.
Depending on this position, the RBC sends the actual maximum   speed allowed ($\vmax$) to the train.  
An obstacle detection sensor sets controller   input $\omega$ to 1 if an obstacle is detected on the track. 
In this situation, the train is expected to brake until it has come to a halt and/or until the obstacle has been removed.    Three position sensors\footnote{For our example, we assume three sensors obtaining location information, for example, from GPS, Balise, and distance radar.}   provide their actual location estimates $x_i \in [x_A,x_B],\ i = 1,2,3$, together with   confidence values $c_i\in [0,1]$. The train controller calculates a fourth location estimate 
based on its last location estimate, last velocity, and last acceleration. We assume that this estimate is associated with a constant confidence value $c_4$.

% ...................................................................................
%% ...................................................................................

The train controller operates in processing cycles of constant time duration $\Delta t$ (a typical value would be $\Delta t = 0.1~{\tt s}$). 
It manages a state tuple   $(x,c,x_4,v,a,\xstop,x_B)$. Here, $x$ denotes the actual aggregated position estimate with its overall confidence value $c$. It is calculated from the sensor values $x_1, x_2, x_3$ and from variable $x_4$ according to Formula~\eqref{eq:weightedsum}. Variable $x_4$ is the current position estimate derived from the physical equation~\eqref{eq:xfour}: $x_4$ estimates the new position based on last position, speed, acceleration, and the time which has passed since the last estimate.  Due to the wheel slip of trains, $x_4$ is only an approximation of the train's true position, and further sensors estimating the position from other sources are required.
 Variable $v$ is the current velocity derived from the physical motion equation~\eqref{eq:vfour}. Output variable $a$ carries the current acceleration value in $\{ a_-,0,a_+\}$. Variable $\xstop$ stores the current estimate where the train would stop, if braking would be started in the next processing cycle. Finally, $x_B$ stores the current movement authority value.   
The initial state is $(x = x_A,c = 1,x_4=x_A,v=0,a=0,\xstop=x_A,x_B=x_A)$. As soon as a movement authority $x_B > x_A+\alpha$ is received, the train controller accelerates the train by setting $a = a_+$. 
The value $\alpha$ is a constant small distance representing the precision  of a train's stopping position: 
if $|x_B - x| \le \alpha$, the train is considered to be ``close enough'' to the position $x_B$ and, therefore, loses its movement authority. When initialising the system, $x_B$ is set to $x_A$, because no movement authority has been received yet.
Conversely, condition $x_B > x_A+\alpha$ indicates that the distance to reach the authorised destination $x_B$ is sufficiently far away from starting point $x_A$, so that the train should be put in motion.
After each processing cycle, the location estimate $x_4$ and the speed estimate $v$ are updated according to the physical equations for motion with constant acceleration:
\begin{align}
\Delta x={}&v\cdot \Delta t+\frac{a}{2} \cdot \Delta t^2,\label{eq:po}\\
\Delta v={}&a\cdot \Delta t,\label{eq:ve}
\end{align}
which imply
\begin{equation}\label{eq:pve}
v+\Delta v=2\cdot\frac{ \Delta x}{ \Delta t} -v.
\end{equation}
%
%Accordingly, the controller assigns new values to $x_4$ and $v$ by
%
%%
%\begin{align}
%v :={}& v + \Delta v \ \text{with $\Delta v$ as defined in \eqref{eq:ve}} \label{eq:vfour} \\
%x_4 :={}& x_4 + v\cdot \Delta t+\frac{a}{2} \cdot \Delta t^2, \label{eq:xfour} 
%\end{align}
%
Accordingly, the controller assigns new values to $x_4$ by
\begin{align}
x_4 :={}& \ul x + v\cdot \Delta t+\frac{a}{2} \cdot \Delta t^2 \label{eq:xfour}, 
\end{align}
where $\ul x$ is the last overall position estimate $x$, calculated at the beginning of the $\Delta t$ period.

%Next, the controller stores the  old overall position estimate $x$ in an auxiliary variable $\ul x$ and
%calculates the new position estimate   $x_4$ according to Equation~\eqref{eq:xfour}, with $\ul x$ replacing  $x$ in this formula.

The overall position estimate is calculated from the current position sensor values $x_1, x_2, x_3$ and $x_4$, taking into account their different confidence values by assigning:
\begin{equation}\label{eq:weightedsum}
x :=\frac{\Sigma_{i=1}^{4} c_i\cdot x_i}{\Sigma_{i=1}^{4} c_i}.
\end{equation}
The new speed estimate is  calculated according to   Equation \eqref{eq:pve} with $\Delta x = x - \ul x$,
and assigned to $v$ as
\begin{align}
v :={}& 2\cdot\frac{x - \ul x}{ \Delta t} - \ul v, \label{eq:vfour}
\end{align}
where $\ul v$ denotes the previous speed estimate.
The overall confidence value for the position value obtained according to Equation~\eqref{eq:weightedsum} is
\begin{equation}\label{eq:confidence}
c = \frac{\sum_{i=1}^4 c_i}{4}.
\end{equation}

If the confidence value is acceptable ($c \ge \cmin$), then the train is accelerated until the maximum speed $\vmax$ has been reached. After that, the train starts coasting. If the confidence value is too low ($c < \cmin$), 
the train is slowed down until it has reached a safe lower speed $\vsafe$ until the position confidence is acceptable again.

With the values for position and speed at hand, the earliest possible stopping position $\xstop$ is calculated under the assumption that the train would 
keep its current acceleration in the actual processing cycle
and  
start braking in the next   cycle. For this calculation, the following assignment can be applied, using an auxiliary variable $\tstop$ indicating the duration until the train comes to a stop.
\begin{align}
\tstop :={}& - \frac{v + a\cdot\Delta t}{a_-}, \label{eq:deltastop}\\
\xstop  :={}& x+v\cdot\Delta t+\frac{a}{2} \cdot\Delta t^2 + (v+a\cdot\Delta t)\cdot\tstop +\frac{a_-}{2}\cdot \tstop^2 \label{eq:xstopa}\\
={}&x+v\cdot\Delta t+\frac{a}{2} \cdot\Delta t^2 -\frac{a_-}{2}\cdot \tstop^2. \label{eq:xstop}
\end{align} 
To understand Assignment \eqref{eq:deltastop}, recall from Equation~\eqref{eq:ve} that the speed changes
to $v' = v + a\cdot \Delta t$ in the current processing cycle, where $a$ is the current acceleration. From the next cycle on, the constant deceleration $a_-$ will be applied, so the speed changes according to $\Delta v = a_-\cdot \tau$ over duration $\tau$. Resolving formula $v' + \Delta v(\tau) = 0$ to $\tau$, yields 
 $\tau = \tstop$ from Assignment \eqref{eq:deltastop} for the duration to come to a halt from velocity $v'$.
For understanding Assignment \eqref{eq:xstop}, recall from Equation~\eqref{eq:po} that the   position
$x$ changes to $x' = v\cdot \Delta t + \frac{a}{2}\cdot\Delta t^2$ in the current processing cycle. After that,
the speed $v'$ has been reached, and deceleration $a_-$ is applied. Applying Equation~\eqref{eq:po} with $v = v'$,   $a  = a_-$, and for the duration $\tstop$, yields the stopping position used in Assignment~\eqref{eq:xstopa}.
Observing that $(v + a\cdot \Delta t) =  -\tstop\cdot a_-$, yields Equation~\eqref{eq:xstop}.

When the estimated stopping position is only $\delta$ meters away from the destination $x_B$, further acceleration is forbidden. As soon as forecast $\xstop$ reaches the value of $x_B$, the train  brakes  until a very low speed $\vmin$ has been reached, from where the train can stop ``immediately'', that is, within less than $\alpha = 0.6~{\tt m}$. If the train has closed in on $x_B$ within $0.6~{\tt m}$, it is braked again until it stops. Then, the train waits for a new movement authority to continue its journey.

Typical values for the constants and boundary variables mentioned in the requirements above are
$$
\begin{array}{rclrclrclrcl}
a_+ & = & 1~{\tt m/s^2},\quad  & a_- & = & -1~{\tt m/s^2},\quad & \Delta t & = & 0.1~{\tt s},\quad & \cmin & = & 0.9 \\
  \vsafe &  =  & 8~{\tt m/s},\quad & \vmax & = & 22~{\tt m/s},\quad&  \vmin & = & 1~{\tt m/s},\quad& \delta & = & 200~{\tt m} \\
  \alpha & = & 0.6~{\tt m} %%%,\quad 
\end{array} 
$$
It can be assumed that $\vmax$ is always greater than $\vsafe$.

% ....................................................................................
\subsubsection*{Formal Controller Model}
The informal system description given above is now modelled using UML state machines~\cite{uml_2_5}. 
The formal model semantics can be specified, for example, by associating a variant of Kripke structures with state machines, as described in~\cite{peleska_sttt_2014}. In the following, we explain the behaviour formalised with these machines in an intuitive way.

The expected controller behaviour is modelled by two state machines. The first updates  the variable vector
$(x,c,x_4,v,a,\xstop,x_B)$ defined above every $\Delta t$ according to the equations listed above. This state machine is not shown, since it consists of a single state with a self loop triggered every $\Delta t$ with an action that just performs the necessary assignments specified above.

\begin{figure}[H]
%%\hspace*{-40mm}
\begin{center}
\includegraphics[width=.8\textwidth]{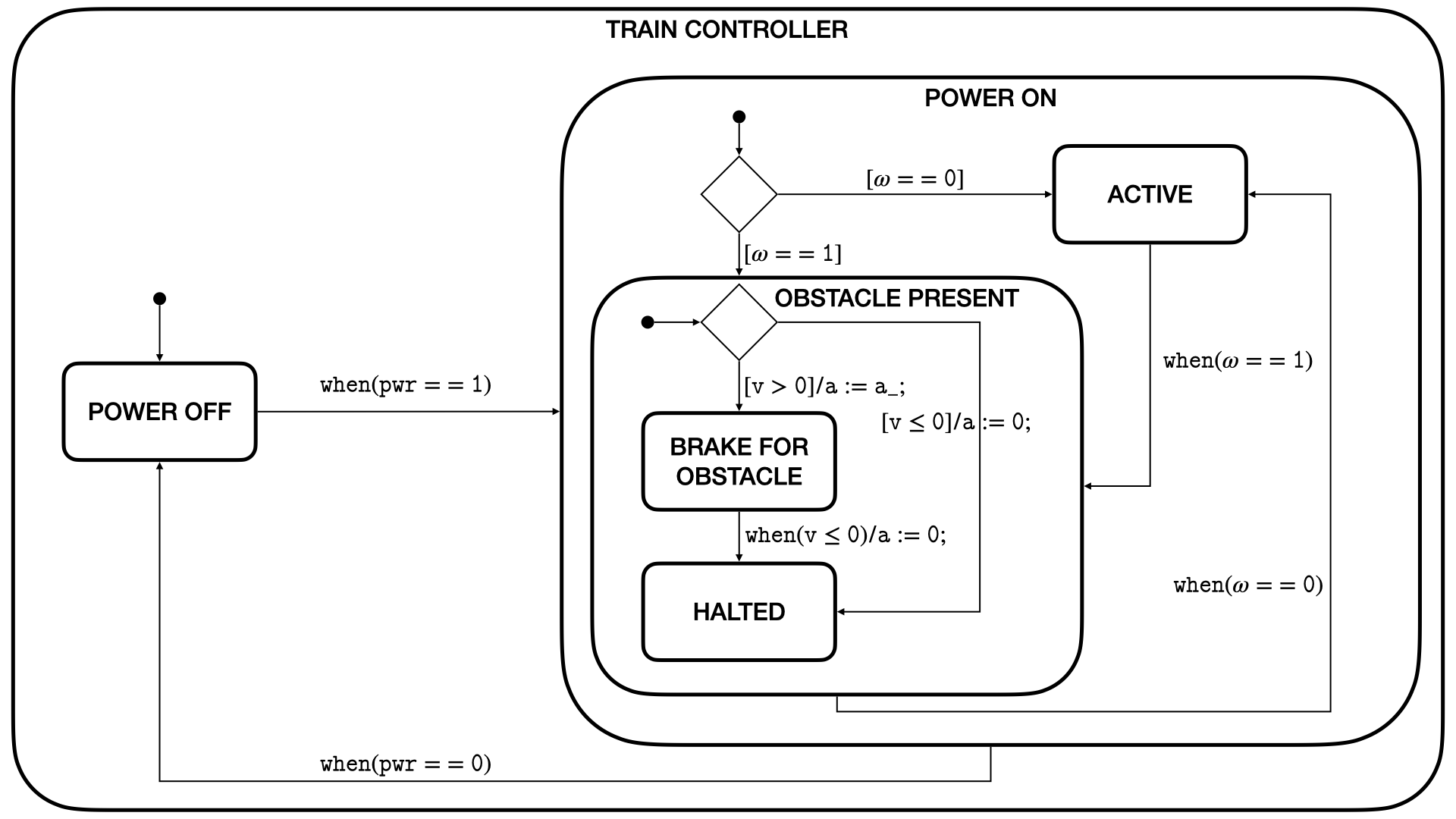}
\end{center}
%%\vspace*{-10mm}
\caption{Train controller behaviour -- top-level state machine.}
\label{fig:behave1}
\end{figure}

\begin{figure}[H]
%%\hspace*{-40mm}
\begin{center}
\includegraphics[width=.7\textwidth]{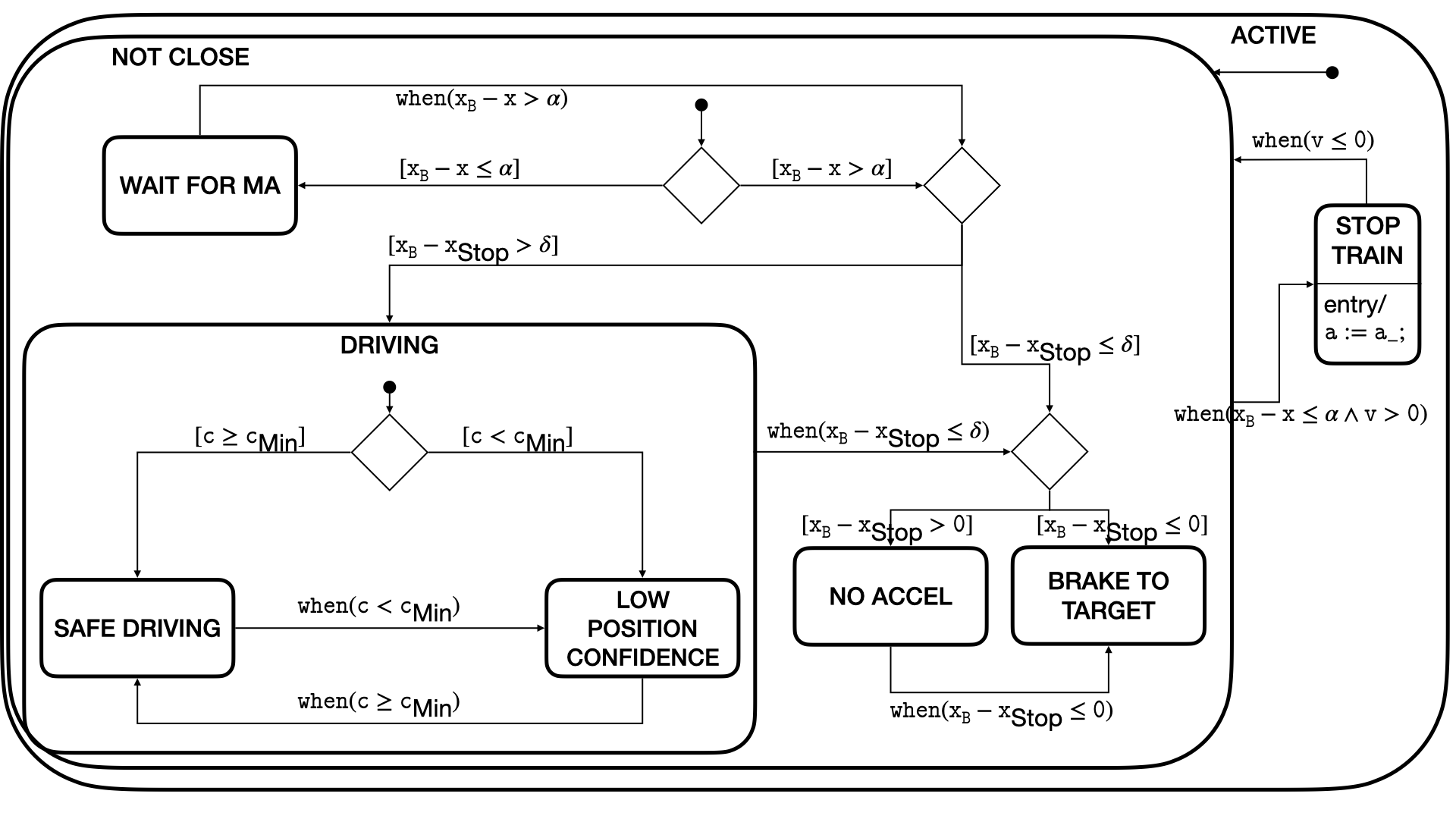}
\end{center}
%%\vspace*{-10mm}
\caption{Train controller behaviour -- active states.}
\label{fig:behave2}
\end{figure}

Concurrently, the hierarchic state machine {\sf TRAIN CONTROLLER} with its top-level machine specified in Fig.~\ref{fig:behave1} is executed. Its normal behaviour is to transit after power on into state {\sf ACTIVE} described 
by the lower-level state machine in Fig.~\ref{fig:behave2}. In the special situation where the controller is 
started with an obstacle in front (condition ${\tt \omega = 1}$), it  transits into  submachine state {\sf OBSTACLE PRESENT}. If the train is still driving, it will brake in state {\sf BRAKE FOR OBSTACLE} until it has come to a standstill and state {\sf HALTED} is entered.
State {\sf ACTIVE}   will always be left   when an obstacle is detected. It is visited again as soon as the obstacle has been removed ($\omega = 0$).

When submachine {\sf ACTIVE} is entered, one or more choice states will be visited to decide which stable state should be chosen. (1) If no movement authority is available ($x_B - x \le \alpha$), state {\sf WAIT FOR MA} is entered.   There, a moving train is braked to stop (this and other subordinate state machine diagrams are not shown here). State 
{\sf WAIT FOR MA} is left as soon as a movement authority is available ($x_B - x > \alpha$).
(2) If a movement authority exists and the train is still far enough from its destination ($x_B - x_\text{Stop} > \delta$), the controller branches into submachine {\sf DRIVING}. There, the train will be accelerated to its
maximum speed $\vmax$. The train will be slowed down if it is too fast and accelerated if it is slower than the maximum speed allowed. If the position confidence value $c$
is too low, the controller transits to state {\sf SAFE DRIVING}, where the train is kept at velocity $\vsafe$ until the position confidence is acceptable again.

% ...................................................................................
%% ...................................................................................

(3) If the predicted stopping location comes as close as $\delta$ to $x_B$ (change condition
$x_B - \xstop \in (0,\delta]$), the train is not allowed to accelerate further. It will be kept at a constant velocity $\vconst\le \vmax$ in state {\sf NO ACCEL} (the associated submachine is not shown). (4) When it is time to brake ($x_B-\xstop \le 0$), state {\sf BRAKE TO TARGET} is entered, where the train
is slowed down to a positive speed value $\vmin$. This positive speed value is maintained until the
train is very close to its destination ($x_B - x \le \alpha$). Then, state {\sf STOP TRAIN} is entered, where the 
train is slowed down to a halt.

% ...................................................................................
\section{Testing on the Module Level -- Complete Methods}\label{sec:module}

% ----------------------------------------------------------------------------------------------
\subsubsection*{Complete Test Suites and Formal Methods}

Since~2011, the  standard RTCA~DO-178C~\cite{DO178C}, which is applicable
for the development and V\&V of safety-critical avionic systems, contains guidelines for the application of formal methods, these are described in the supplement RTCA~DO-333~\cite{DO333}. This supplement lists model checking as one accepted formal verification method among others (e.g.~theorem proving). The complete testing strategies advocated in this paper for module-level software verification can be regarded as model checking of code in two variants: (1) for \emph{conformance checking}, \emph{complete test suites} proving model conformance can be used.
(2) For \emph{property checking}, requirements for the given module need to be formalised (typically in a temporal logic, e.g.~LTL). Then it can be verified again by means of complete test suites that the code fulfils the given formula, just as the reference model does. Completeness of a test suite asserts that every violation of model conformance or of the specified property will be uncovered, and that correct implementations will pass the test suite. Frequently, it suffices to apply \emph{exhaustive} test suites. These guarantee to detect errors but may also reject correct implementations in certain situations. 

In black-box testing, completeness or exhaustiveness can only be guaranteed under certain hypotheses, typically about the maximum number of states implemented in the SUT, and about  
the possible mutations of branching conditions and output expressions. For safety-critical systems development, however, software source code must be available for analysis. Therefore, the hypotheses about the SUT can be checked by means of various forms of static analysis. Consequently, complete or exhaustive test suites correspond to ``real'' code-level model checkers in the sense that the absence of failed test executions {\it proves} software correctness. Different from model checkers, the verification is performed by dynamic execution of the test cases against the code, and not by static code analysis methods.

The autonomous train controller example introduced in Section~\ref{sec:train} has typical characteristics occurring in many ATV control applications: inputs may be associated with conceptually infinite domains, such as real values for train position, but outputs are discrete, as the accelerate/coast/brake outputs to the train engine.\footnote{Other examples of systems of this kind are airbag controllers, thrust reversal controllers in aircraft, or safety controllers for robots interacting with humans.}
For this class of systems, complete test suites can be constructed in four steps~\cite{peleska_sttt_2014,Huang2017}. 

\begin{enumerate}
\item[Step~1] The input equivalence classes are constructed. 

\item[Step~2] The system model is abstracted to a finite state machine (FSM) in Mealy style. The inputs of this FSM correspond to the input equivalence classes of the original system  model, and the FSM outputs correspond to the discrete system outputs. 

\item[Step~3] A complete FSM test strategy is applied, resulting in a test suite with input sequences of equivalence class identifiers and expected outputs from
the domain of system outputs. 

\item[Step~4] This FSM test suite is translated into a test suite for the original system, by selecting concrete representatives for each of the classes referenced in an FSM test case. The resulting test suite has the analogous completeness properties as the FSM test suite (this is the main result shown in~\cite{peleska_sttt_2014,Huang2017}).

\end{enumerate}

\begin{example}{ex:classes}
For the autonomous train controller, we consider two modules: the cyclic update machine $C_0$ providing new position and speed values on every $\Delta t$ cycle, and the proper controller $C_1$ exercising the acceleration commands to the engine according to the SysML state machines shown in Fig.~\ref{fig:behave1} and subsequent sub-machine models.

For   controller module $C_1$, the SysML state machine model is associated with a formal behavioural model 
by means of a translation to a
flattened
\emph{symbolic finite state machine (SFSM)}~\cite{DBLP:conf/icst/Petrenko16,krafczyk_niklas_2021_5151778}. These can be considered as a simpler subset of the more general Kripke structures used in~\cite{peleska_sttt_2014,Huang2017} for
the elaboration of the complete testing theory. The SFSM representation has the advantage that the input equivalence classes can be easily calculated by enumerating all positive and negated conjunctions of the SFSM guards, as long as these have a model.  For the train controller module $C_1$ shown in Fig.~\ref{fig:behave1}  and its sub-machines, this results in 28 input equivalence classes $c_i,\ i=1,\dots,28$, of which we show several examples here:
\begin{eqnarray*}
c_1 & \equiv & {\tt pwr} = 0 \\
c_3 & \equiv & {\tt pwr} = 1 \wedge \omega = 1 \wedge 0 < v\\
% i05:pwr=1&w=0&x_B-x>alpha&x_B-x_Stop>delta&c>=c_Min&0<v<v_Min 
c_5 & \equiv & {\tt pwr} = 1 \wedge \omega = 0 \wedge x_B - x > \alpha \wedge x_B - \xstop > \delta 
\wedge c\ge \cmin \wedge 0 < v < \vmin \\
c_9 & \equiv & {\tt pwr} = 1 \wedge \omega = 0 \wedge x_B - x > \alpha \wedge x_B - \xstop > \delta 
\wedge c\ge \cmin \wedge v > \vmax \\
% i27:pwr=1&w=0&x_B-x<=alpha&x_B-x_Stop<=0&v<=0  
c_{27}  & \equiv & {\tt pwr} = 1 \wedge \omega = 0 \wedge x_B - x \le \alpha \wedge x_B - \xstop \le 0 \wedge v = 0
\end{eqnarray*}
Input equivalence class $c_1$ is specified by all input tuples where ${\tt pwr} = 0$; all other inputs can have arbitrary values. This holds because the other inputs do not affect the behaviour of $C_1$, as long as the power is off. In contrast to this, the predicate specifying class $c_9$ consists of
six conjuncts involving all   inputs to $C_1$. It captures the situation where the train is still sufficiently far away from the target station and the position confidence is high, so that it may drive with maximum speed, but the train is currently overspeeding. Note that all conjuncts come from (possibly negated) guard conditions in the state machine.

Applying the abstraction principle described in~\cite{peleska_sttt_2014,Huang2017} results in an FSM
with input alphabet $c_1,\dots,c_{28}$, output alphabet  $a := a_0, a := a_-, a := a_+$, and six states.
To this FSM, we apply  the well-known complete H-Method~\cite{DBLP:conf/forte/DorofeevaEY05}. This method produces test suites guaranteed to uncover all violations of observational equivalence, provided that an upper bound of the internal number of states in the implementation is known. Since we assume that the source code of $C_2$ is available (this would always be required for safety-critical control components), 
the number of control states used in the $C_2$ implementation can be extracted by means of static analysis.
If the code is directly generated from the SFSM, it is guaranteed that the implementation also has only
six non-equivalent control states. Under this assumption, application of the H-method 
 results in approximately~600 test cases\footnote{The number of test cases needed usually varies with the method implementation. We have used here the open source library {\tt libfsmtest} 
for model-based testing with FSMs, which is available from \url{https://bitbucket.org/JanPeleska/libfsmtest}.}.
As an example of these generated test cases, consider the following one, consisting of two steps.
% i27:pwr=1&w=0&x_B-x<=alpha&x_B-x_Stop<=0&v<=0/a0 [PO_BFO_A.WMA], 
% i05:pwr=1&w=0&x_B-x>alpha&x_B-x_Stop>delta&c>=c_Min&0<v<v_Min/a+ [A.D]
\[
    (c_{27}/{\tt a := 0}).(c_5/{\tt a := a_+})
\]
Translation of this FSM test case to a concrete 
test case, which can be executed against the target system, requires to select representatives from input classes $c_{27}$ and $c_5$   specified above. An example for such a concrete test case is (we assume that $x_A = 0$)
\[
\begin{array}{l}
    ({\tt pwr} = 1 \wedge \omega = 0 \wedge x_B = 0 \wedge x = 0 \wedge \xstop = 0 \wedge v=0/{\tt a := 0}). \\
   \hspace*{15mm}  ({\tt pwr} = 1 \wedge \omega = 0 \wedge x_B = 10000 \wedge  x = 0 \wedge   \xstop = 0 
\wedge c=0.95 \wedge v = 0.01/{\tt a := a_+}).
\end{array}
\]
In the first step, the train controller is powered, and no obstacle is present. The speed of the train is zero, and there is no movement authority available ($x_B = 0$). Thus, it is expected that the requested acceleration is zero, i.e.\ ${\tt a := 0 }$. In the second test step,
a movement authority arrives ($x_B = 10000$), so it is expected that the train starts to accelerate, i.e.\ ${\tt a := a_+}$.

The main results of~\cite{peleska_sttt_2014,Huang2017} imply that the   test suite, where all test cases of the abstract FSM suite are made concrete as illustrated in the example case above, is also complete under certain alternative sets of sufficient conditions. The simplest   condition set requires that the implementation does not introduce additional states, and   that the implementation uses the same guard conditions as the SFSM. Both conditions can be checked by means of static analysis. Note that, if these conditions apply, it always suffices to select a single representative from each input equivalence class $c$, and   use this representative for all concrete test cases where $c$ is referenced in a test step.

The longest test cases of this complete test suite have only four steps: every state of the minimised FSM can be reached in at most two steps, when starting from the initial state.  From every state, every input needs to be exercised; this adds one further step to the test case. Then, the target state reached needs to be distinguished from other states. This can be performed for the train controller by traces of
length one. This adds up to test cases of maximum length four.
\end{example}

It has been criticised in the past that complete test suites turn out to be too big to be applied in practice to systems of realistic complexity. We agree that complete test suites should not be applied to system level testing. However, for software components of a complex system, these test suites turn out to be feasible from today's perspective, for the following reasons: (1) As shown in~\cite{Huebner2019,DBLP:conf/rssrail/PeleskaHH16}, the construction of equivalence classes reduces the test suite size significantly without impairing the suite's completeness properties. (2) For the application of complete test suites as described here, we need not formulate hypotheses about potential additional control states in the implemented software, because the number of these states can be extracted from  the code by static analysis. Consequently, the test suite size only depends polynomially on the number of states~\cite{chow:wmethod} (the size would grow exponentially in the potential 
number of {\it additional} control states present in the implementation). (3) The execution of software tests can be performed
fully automatically in the cloud, so that many test cases can be executed in parallel. For these reasons, the original criticism of complete test suites no longer applies today.

In summary, the complete test suites advocated here are suitable 
for verifying the logical correctness of the source code, but they are not intended for receiving certification credits regarding the correctness of the integrated hardware/software (HW/SW) system.
In the approach advocated here, this integration aspect is shifted to the system test level.

%%% model -- synthesiser --- code

% ----------------------------------------------------------------------------------------------
\subsubsection*{Model-Conformance vs.~Requirements Satisfaction -- Traceability Issues}

The supplement to RTCA~DO-178C on model-based development and V\&V states~\cite[MB.6.7]{DO331}:
{\sl ``Model coverage analysis does not eliminate the need for traceability analysis between requirements from which the model was developed and the model.''} This has an implication on the model-based testing approach which -- to the best of our knowledge -- has not received sufficient attention in the research communities: traceability demands that requirements are related to the model portions realising these requirements. Then, the code has to be related to the model parts it implements. Finally, the test cases checking the code need to trace back to the requirements they intend to verify.\footnote{This can be done either indirectly through the traceability chain {\sl test case $\longrightarrow$ code $\longrightarrow$ model $\longrightarrow$ requirement} or directly from test case to requirement.} It does {\it not} suffice to argue that the formal model
(in our case, the SysML state machines) has been reviewed (or even formally verified) and found to reflect all requirements correctly and then conclude that, since the code has been shown by complete test suites to be equivalent to the model, 
it must implement all requirements ``somehow'' in the proper way.

For modelling with SysML, traceability between model elements and requirements can be directly represented by means of the 
so-called {\sf \guillemotleft satisfy\guillemotright} relationship linking behavioural or structural model elements to requirements~\cite{SysML15}. It has been shown in~\cite{DBLP:conf/isola/0001BH18} how this information can be exploited to create formal representations of requirements in a temporal logic like LTL. The atomic propositions of the LTL formulae 
are quantifier-free first-order expressions over model variables, including inputs and outputs. The validity of a formula can be decided by abstracting model computations to the sequences of sets containing the formula's atomic propositions valid in the respective computation 
step, see Section~3.2.2 of~\cite{DBLP:books/daglib/0020348}. In~\cite{krafczyk_niklas_2021_5151778}, it has been shown that for given sets $P$ of atomic propositions exhaustive test suites can be constructed, which frequently require fewer test cases than needed for establishing full model equivalence. Exhaustiveness in this situation means that the test suite is guaranteed to fail for at least one test case, if the computations of the implementation, when abstracted to traces over sets of elements from $P$, contain traces that are not produced by the reference model. Therefore, passing the test suite implies that the implementation fulfils all LTL properties over atomic propositions from $P$ that are also fulfilled by the reference model.

We conclude that the requirements-based verification tasks imposed by the RTCA~DO-178C standard can also be handled by means of exhaustive test suites. We suggest that on the module level two complete test suites should be executed: the first, with the objective to prove model equivalence, enables us to calculate a meaningful test coverage measure for system tests, as will be explained in the
next section. The second, requirements-driven test suite, is essential to satisfy the traceability requirements of the standard.

% ==============================================================================================
\section{System-Level Testing}\label{sec:systemtest}

% ----------------------------------------------------------------------------------------------
\subsubsection*{Meaningful System Tests}

Besides checking the correctness of the overall system integration,
a major objective of system testing is to verify the complete workflow of services and applications, potentially across several communicating  controllers and the respective interfaces involved. Tests designed for this purpose are usually called 
\emph{end-to-end (E2E) tests}. Analysing the module tests discussed in the previous section, it becomes apparent that they do {\it not} represent meaningful E2E tests. This is illustrated by the following example.
\begin{example}{ex:moduletestsnogoodforendtoend}
%i04:pwr=1&w=0&x_B-x>alpha&x_B-x_Stop>delta&c>=c_Min&v<=0/a+ [A.D], 
%i28:pwr=1&w=0&x_B-x<=alpha&0<v/a- [A.ST], 
%i04:pwr=1&w=0&x_B-x>alpha&x_B-x_Stop>delta&c>=c_Min&v<=0/a+ [A.D], 
%i06:pwr=1&w=0&x_B-x>alpha&x_B-x_Stop>delta&c>=c_Min&v=v_Min/a+ [A.D]
Consider one of the module tests of maximum length four from the complete test suite discussed in Section~\ref{sec:module}. We present here the symbolic version with input equivalence classes, and not the concrete version, where atomic propositions have been resolved by selecting concrete values. 
\[
\begin{array}{ll}
\text{Step~1.} &     ({\tt pwr} = 1 \wedge \omega = 0 \wedge x_B -x > \alpha \wedge x_B-\xstop > \delta\wedge c\ge \cmin 
    \wedge v=0/{\tt a := a_+}). \\
\text{Step~2.} &    ({\tt pwr} = 1 \wedge \omega = 0 \wedge x_B -x \le \alpha   
    \wedge 0< v/{\tt a := a_-}). \\
\text{Step~3.} &    ({\tt pwr} = 1 \wedge \omega = 0 \wedge x_B -x > \alpha   \wedge x_B-\xstop > \delta \wedge c\ge \cmin
    \wedge v=0/{\tt a := a_+}). \\
\text{Step~4.} &   ({\tt pwr} = 1 \wedge \omega = 0 \wedge x_B -x > \alpha   \wedge x_B-\xstop > \delta \wedge c\ge \cmin
    \wedge v = \vmin/{\tt a := a_+})    
\end{array}
\]
The first test step initiates a power up in a situation where movement authority is already available ($x_B-x>\alpha$).
The train is expected to accelerate.
Step~2, however, initiates a robustness transition in the controller, because the train location appears to be close to the destination ($x_B-x\le\alpha$), without having first fulfilled condition $x_B-\xstop \le \delta$ which would be expected in a ``physically reasonable'' execution. The train is expected to brake. While this transition is contained in the controller model to ensure its robustness, its occurrence is highly unlikely, so that it would not be selected for a ``realistic'' E2E test,  but only for   a robustness test. It would suffice, however, to check this aspect of robustness at the module level. 
In Step~3, a situation is described again where the train is further away from its destination -- this is expressed by 
conditions $x_B -x > \alpha   \wedge x_B-\xstop > \delta$. If we just selected a smaller value of $x$ to fulfil these conditions, this would again result in a robustness transition: the position sensors provide an improved position value, while the previous one for Step~2 was too high. Alternatively, this step could be realised by increasing the value of $x_B$. This would correspond to a realistic system test step, where a new movement authority is obtained by increasing the destination value $x_B$. Obviously, the distinction between a robustness test step and a normal behaviour test step suitable for an E2E test cannot be made on the basis of the equivalence class formulae alone. While this distinction is not relevant for module testing, it has high significance for the design of meaningful E2E tests.
In the last step, the input is just updated according to the expected change of velocity: if the speed was zero before and the train keeps accelerating, the velocity will increase to $\vmin$.

The test case now ends while the train is still accelerating. For a suitable E2E test, we would expect a continuation with further test steps, leading the train to its destination $x_B$. Such continuations, however, do not exist in the module test suite, because every test case there starts from the initial controller state by powering the system.

Obviously, the physical model connecting acceleration, velocity, and position (see Section~\ref{sec:train}) needs to be considered for constructing meaningful system tests in a simulation environment: Step~3, for example, can only be executed when the train has come to a standstill ($v=0$), and the point in time when this happens depends on the effect of the negative acceleration $a_-$ and the actual speed value when the brakes had been triggered in Step~2.

For system tests on target hardware and in the real operational environment, these simulations of the physical world are unnecessary, but the trigger conditions for each new test step need to be monitored. Here, Step~3 can only be triggered {\it after having observed} that the train has come to a standstill. Then, the new value of $x_B$ can be provided to the SUT, so that $x_B -x > \alpha   \wedge x_B-\xstop > \delta$ is fulfilled. Moreover, we observe that
system testing with the target system and its operational environment needs to be {\it intrusive}, if different 
confidence values of the position sensors and erroneous position values need to be tested. `Intrusive' means that   the behaviour of the position sensors can be changed for test purposes, for example, by corrupting position values and 
lowering the actual confidence values calculated by the sensors. Finally, the occurrence and removal of obstacles has to be controlled (for example, remotely controlled vehicles could be moved on the track and removed at a later point in time). 
\end{example}

Summarising the findings highlighted by this example, we conclude that
(1) Module test suites are not well-suited to be turned into meaningful E2E tests.
(2) The formula representations of input equivalence classes do not provide sufficient information about which concrete solutions would give meaningful test data for E2E testing.
(3) Test environments for the simulated execution of system tests require components observing the applicable physical laws during test execution.
(4) System tests with the ``real'' target in its operational environment require intrusive test equipment for some input interfaces to the SUT.

Finding number one   can be justified more formally. When testing for model conformance, the complete strategies applied usually strive to reach every state of the SUT on the shortest path possible, and then to execute every possible input from this state. Consequently, breadth-first-search algorithms are applied.
This has the effect that   transitions that should   only be performed in exceptional cases  are frequently taken if they reside on the shortest path to  
such a state. Moreover, the objective to exercise every possible input (or input class) in a given state necessarily leads to frequent resets, leading to the start of a new test case. In contrast to this, E2E tests need to be identified by  a depth-first search through a system model; this leads to longer test cases starting and ending in system states that are meaningful from the perspective of the tested application.

As a consequence of these observations, we see that it is generally impossible to ``glue'' several module tests together and come out with a meaningful E2E test. We further observe that the optimal error detection capabilities of complete test suites are unrelated
to their end-to-end test quality.

% ....................................................................................
\subsubsection*{System Test Scenarios}

System tests of autonomous trains need to be executed in different \emph{scenarios}, where each scenario is specified by a route through the rail network and a set of events (low confidence positions, obstacles, overspeeding) to occur at certain points while traversing the route. This concept is similar to scenario-based testing of autonomous road vehicles~\cite{hungar_scenario-based_2018}, but considerably simpler, since the train movements are restricted by the rail network, and the absence of collision hazards involving 
other trains is already ensured by the interlocking system.

\begin{figure}[h]
%%\hspace*{-40mm}
\begin{center}
\includegraphics[width=\textwidth]{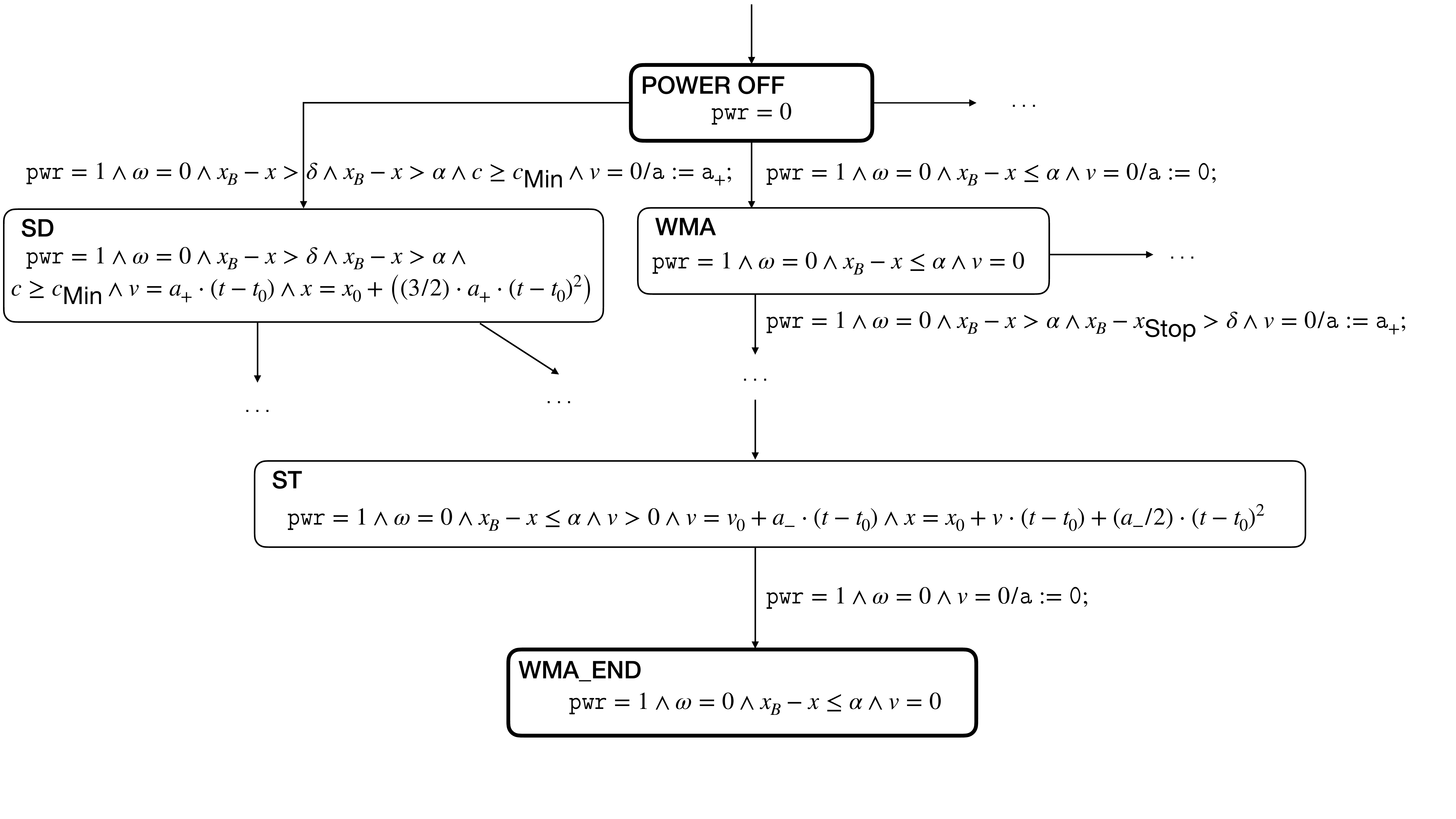}
\end{center}
%%\vspace*{-10mm}
\caption{Partial representation of a symbolic scenario test tree (SSTT) for the autonomous train control system.}
\label{fig:sstt}
\end{figure}

To allow for automated test case derivation for E2E system tests, we introduce a data structure called
\emph{symbolic scenario test tree (SSTT)}.
These trees are interpreted as  \emph{hybrid systems}~\cite{Hen96} which are simplified in the sense that our outputs are always concrete assignments instead of being  implicitly specified through jump conditions. The use of these trees is exemplified 
with the tree fragment shown in Fig.~\ref{fig:sstt}.

% ...................................................................................
%% ...................................................................................

Just as the control modes of hybrid system models, 
the nodes of an SSTT represent invariants over both discrete and time-continuous variables.
The invariants involving discrete variables can be directly derived from the state machine models created for the
controller modules. The time-continuous evolution of physical inputs, such as location and velocity in relationship to acceleration and passing time, need to be derived from the physical laws applicable in the system environment. The time-dependent invariants shown in the states of Fig.~\ref{fig:sstt} are all
implied by the physical laws presented in Section~\ref{sec:train}.

The edges of an SSTT are labelled with guard conditions and output assignments, as in symbolic finite  state machines. In Fig.~\ref{fig:sstt}, the root of the tree is represented by node {\sf POWER OFF}, 
which only has the invariant ${\tt pwr} = 0$. From there, edges lead to different child nodes, depending on the guard conditions labelling these edges. The edge to node {\sf WMA} is taken when the power is switched on, no obstacle is present, the train is without movement authority ($x_B-x\le \alpha$), and it is not moving. Therefore, the acceleration can be set to 0, and the system resides in state {\sf WMA} until a movement authority arrives. There are several situations to distinguish: the movement authority might arrive with an obstacle present, or it might arrive in a   position that is already quite close to the destination. The downward edge shown in Fig.~\ref{fig:sstt} specifies the
situation where no obstacle is present, and the train is not yet required to brake for the
target destination ($x_B - \xstop>\delta$).
The edge from node {\sf POWER OFF} to node {\sf SD} specifies the situation where the train is powered, while a movement authority is already available, so that the train is allowed to accelerate, directly leaving its standstill position.
If the invariants need to refer to variable values at node entry, these are denoted by ${\tt <variable>_0}$. 
In node 
{\sf SD}, for example, $(t-t_0)$ denotes the time which has passed since entering the node.
Towards the leaves of the tree, the nodes describe the situations where the train approaches the destination $x_B$ and brakes until it comes to a standstill. This is exemplified here by nodes {\sf ST} (``stop train'') and {\sf WMA\_END}
(``wait for next movement authority'').

As can be seen in this example, the SSTT contains the necessary information about physical laws to be applied. These were missing in the module test models. Moreover, the SSTT is created in such a way that robustness transition sequences that are better tested on the module level are left out, and that the leaves of the tree represent suitable target states from the perspective of E2E testing. The example also indicates that these trees can become quite large. In part, they can be automatically derived from 
the state machines specified for the controller modules in combination with rules about applicable control laws. The 
  information about sequences of transitions that are only taken in robustness situations, however, need to be provided by system experts. Moreover, the suitable termination states for E2E testing need to be manually provided. In~\cite{DBLP:conf/isola/0001BH18}, methods for automated elaboration of requirements formulae from SysML models have been presented. With these (LTL-) formulae at hand, it is at least possible to check
automatically whether the SSTT   covers a certain requirement. This, however, does not yet imply that the respective path through the tree also represents a meaningful E2E test. For these reasons, we expect that SSTT creation can only be partially automated and will always require inputs from   experts.

% ----------------------------------------------------------------------------------------------
\subsubsection*{Cloud-based Tests vs.~Tests on the Target System -- System Test Coverage}

Cloud-based testing significantly increases the availability of hardware resources for test campaigns, and 
improves the possibilities to extend or shrink these resources according to the demands of such campaigns.
It has to be discussed, however, how certification credit may be obtained for tests performed in the cloud, since cloud platforms do not allow for execution of SUTs with their original hardware.
The standard supplement RTCA~DO-333 regulating the use of formal methods~\cite[FM.6.3.1.c]{DO333}
states that formal verification results can also verify aspects of compatibility of software and hardware, if the   underlying formal models also cover the hardware platform. For specific verification objectives like worst-case execution time analysis, this possibility has already been successfully exploited~\cite{DBLP:journals/tecs/WilhelmEEHTWBFHMMPPSS08}.
For system testing in the cloud, more general hardware models are needed, so that control applications can be executed on these models with their original machine code and address maps, as used on the real target. Such models have recently been developed; they are called \emph{virtual prototypes}~\cite{DBLP:journals/jsa/HerdtGPD20} and represent more advanced variants of the simulators that have been used for quite some time. We expect that certification credit for cloud-based   system tests or HW/SW integration tests can be achieved if the software is executed in such a variant of virtual prototypes. This option, however, requires a new version of the standards and would certainly require \emph{tool qualification}~\cite{DO330} for the virtual prototypes, to justify that they really represent a faithful model of the target HW. 

For good reasons, we can also expect that in the future at least some tests will be required to be performed on the ``real'' target system with its integrated hardware and software:
RTCA~DO-333   states~\cite[FM.6.7]{DO333}
{\it ``Tests executed in target hardware are always required to ensure that the software in the target computer will satisfy the high-level requirements \dots''}.
We expect that system-level tests performed on the target system will be  regarded as more valuable than module tests on target HW, since system tests exercise more aspects of drivers and firmware than module tests that often only cover software interfaces and do not stimulate drivers and firmware at all.
Since the current standard ISO~26262 applicable in the automotive domain refers to RTCA~DO-178 when suggesting V\&V methods, we expect that the verification approaches accepted for the avionic domain will also remain admissible for the automotive domain in the future. In the railway domain, the applicable standard 
EN~50128~\cite{CENELEC50128} also regulates (even requires) the use of formal methods, but is less explicit than RTCA~DO-178C with respect to formal hardware models and the amount of testing to be performed on the target system.
With these standard-related facts in mind, we conclude that the most specific guidelines for a formal approach to testing in simulated and in target environments are given by the avionic standards.

According to these   standards,   system requirements shall all be covered by system-level tests. As discussed in Section~\ref{sec:module}, requirements can be transformed into LTL formulae, and the symbolic scenario test tree introduced earlier contains sufficient information about valid atomic propositions along each path to decide with an automated procedure which formulae are fulfilled on each path.\footnote{Note that this procedure requires 
a \emph{vacuity analysis}~\cite{ball_vacuity_2008}: a requirement of the form
$\tg (\varphi\Rightarrow\psi)$ may be satisfied on a path $\pi$ of the tree just because $\varphi$ never becomes true. In this case, the path $\pi$ would not be a suitable witness for testing this requirement.}
Uncovered requirements may both lead to new branches to be added into a scenario test tree and to the creation of new scenarios, if the existing ones cannot accommodate paths where a given formula can be checked non-vacuously.

The new aspect not covered by current standards is the fact that a major portion of the system tests will be executed in the cloud, and not on the target within its operational environment. We expect that future revisions of the standards will demand that ``a reasonable portion'' of these tests should be executed on the target. Then, the question that remains to be answered is which sub-collection of system tests represents such a reasonable portion. When the approach advocated in this paper is followed, a sound answer to this question can be given: system testing on the target should cover the normal behaviour transitions of the SFSM models used for module-level testing. This is justified since we have shown that the software conforms to its SFSM models, so SFSM transition coverage implies software transition coverage. Consequently, transition coverage of SFSMs implies that all relevant transitions of the software have been exercised {\it on the target system}. This is a very comprehensive check of the HW/SW integration correctness. We suggest to exempt robustness transitions from this coverage requirement, since many robustness aspects can be better tested on the module level, avoiding too many variants of intrusive tests on the system level. Another suggestion is to let the required degree of transition coverage depend on the \emph{design assurance level (A --- D)}, that is, the SUT criticality. This concept is currently adopted in~\cite{DO178C} for the code coverage required to get certification credit.

% ----------------------------------------------------------------------------------------------
\subsubsection*{Agent-based Approach and Online Testing}

The preceding discussion of system tests suggests a testing environment where multiple concurrent components are deployed that are responsible for the different aspects identified above.
(1) Coverage analysis and coordination of concurrent test executions.
(2) Control of an individual system test execution.
(3) Simulation of the physical environment in real time or simulated time (for cloud-based system tests). 
(4) Interfacing to the target system in its operational environment (transmission of movement authorities, intrusive influence of the position sensors, and obstacle creation and removal).
  
For realising these testing environments, we  apply  agent-based system testing (ABST).   
To this end,  a multi-agent system (MAS) is realised according to the paradigm of \emph{mixed initiative control}. As stated in~\cite{10.1145/1514095.1514126},
this form of agent collaboration is characterised by {\sl ``allowing a supervisor and group of searchers to jointly decide the correct level of autonomy for a given situation''}. In~\cite{10.1145/1514095.1514126}, this paradigm was 
intended to regulate the collaboration between multiple robots and a human operator. In our context here, a coordinator test agent can monitor the transition coverage achieved so far and  delegate the respective test steps suitable for covering missing transitions to  agents executing tests on the target and its operational environment. Two further classes of test agents simulate the physical environment and interface to the SUT hardware for tests on the target.
It has  already been shown in~\cite{DBLP:conf/aitest/ChanceGLPE20}, how MAS technology helps to increase the test strength of system test cases by letting agents execute strategies for selecting test data which is most effective in the current situation of a system test execution.  

In~\cite{DBLP:conf/fates/LarsenMN04}, the authors introduced the term \emph{online testing} 
(or \emph{on-the-fly testing}) for model-based tests combining the generation of test  steps
of a test case with the actual execution and checking of expected results.\footnote{This concept has already been 
used in 1996 in the VVT-RT Tool~\cite{DBLP:conf/fm/Peleska96} whose commercial version is called RT-Tester today \url{www.verified.de}.} Online testing is very well-suited for the objectives outlined in this paper, because in the presence of potentially millions of system test cases to be executed, a separate test generation phase preceding the test execution might take too much time, whereas online testing delivers information about passed and failed test steps right from the start.

Moreover, online testing helps to let the SSTT grow incrementally, instead of creating the complete tree beforehand. The coordinator test agent can initialise the SSTT with just as many different paths as test execution agents are available. Monitoring the module transition coverage achieved with these system test scenarios,  the agent can let the SSTT ``grow'' by adding new paths that are suitable to cover just the unvisited module transitions.

% ==============================================================================================
\section{Conclusion and Future Work}\label{sec:conc}

In this paper, we have proposed a novel approach for efficient module testing and system testing of autonomous transportation systems (vehicles, trains, aircraft). Techniques of this kind are of considerable importance, because the number of tests to be performed on autonomous systems  is so much higher in comparison to test suites for conventional systems. Therefore, these test campaigns cannot be performed in acceptable time with conventional methods, since the latter  impose system tests to be executed with original equipment only. The key characteristics of the approach proposed here are (1) the use of complete test strategies on the module level, (2) concurrent online systems tests executed  on original equipment and in cloud-based simulation environments, and (3) the use of test agents to coordinate the system test effort. The consistency of this approach with applicable standards has been explained.

Two important aspects of testing autonomous systems were beyond the scope of this paper. The first is the fact that
many autonomous systems are so complex that they can no longer be represented by comprehensive models. Instead, they are specified in scenario libraries, and this leads to the questions of scenario completeness and consistency~\cite{DBLP:conf/itsc/HauerSHP19,DBLP:conf/isola/000120}. The second aspect concerns the assurance 
to be provided for applications based on neural networks and multi-agent systems, because their behaviour cannot be specified and verified with conventional methods, and their behaviour may evolve over time, due to machine learning effects and on-the-fly modification or even creation of plans~\cite{DBLP:conf/eccv/SunCHK20}.

Future work will focus on the detailed evaluation of the approach advocated in this paper. At first, an in-depth  analysis   for the autonomous train control system presented here will be performed. Next, applications from the 
robotic domain (human-robot collaboration) and the 
avionic domain, such as autonomous taxiing, take-off, and landing or formation flight with the objective of fuel saving will be investigated. 
Future work will also cover distributed autonomous systems under test, based on existing results such 
as~\cite{DBLP:journals/cj/Hierons16,DBLP:journals/access/LimaFH20}.

% ==============================================================================================
\bibliographystyle{eptcs}
\bibliography{references,jp,hidyve}
\end{document}